\documentstyle[12pt]{article}
\def\be{\begin{eqnarray}}
\def\ee{\end{eqnarray}}
\def\ba{\begin{array}}
\def\ea{\end{array}}
\begin{document}
\begin{center}
{\Large
{\bf Generation of the bosonic string theory solutions from\\
\vskip 5mm
     the stationary Einstein fields via projection symmetry}}
\end{center}
\vskip 15mm
\begin{center}
{\bf \large {Oleg V. Kechkin}}
\end{center}
\vskip 15mm
\begin{center}
DEPNI, Institute of Nuclear Physics,\\
M.V. Lomonosov Moscow State University,\\
Vorob'jovy Gory, \, 119899 Moscow,  Russia,\\
e-mail: \,\,\,\, kechkin@depni.npi.msu.su
\end{center}
\vskip 15mm
\begin{abstract}
A new formalism for generation of solutions in the consistently truncated
low-energy bosonic string theory is developed. This formalism gives
a correspondence of the projection type between the theories toroidally
compactified from the diverse to three dimensions. Taking the stationary
Einstein-dilaton gravity as the theory with a lower dimension, we generate
its bosonic string theory extension and calculate the bosonic string theory
solution corresponding to the Kerr-NUT one modified by the presence of the
charged dilaton field.
\end{abstract}


\newpage
\renewcommand{\theequation}{\thesection.\arabic{equation}}
\section{Introduction}
At low energies the bosonic string theory becomes a field theory
of its massless modes \cite{Kir}. These modes include the dilaton
and Kalb-Ramond fields coupled to gravity. The effective gravity
model lives in the multidimensional space-time and allows a
toroidal compactification \cite{MahSch}. The resulting dynamical
system possesses a rich group of hidden symmetries; in the case
of the compactification to three dimensions this group coincides with
$O(d,d)$, where $d$ is the number of the compactified dimensions
\cite{Sen3}. In fact $O(d,d)$ is the isometry group for the
$O(d,d)/O(d)\times O(d)$ symmetric space model arising in this
situation. The isometry transformations act in the solution space
and give a base for the solution generation technique in the theory
under consideration (see \cite{EMT}
for the Einstein and Einstein-Maxwell theory analogies and \cite{Youm}
for the string theory applications).

In this article some special truncation of the bosonic string theory
is considered \cite{HK}; the truncated theory is compactified to three
dimensions on a torus. In fact such a procedure gives a series of
the resulting 3-dimensional symmetric space models coupled to the
effective 3-gravity: a concrete representative of this series is
labeled by the number $d$. Our main goal is to establish a general map
which relates the models with different numbers of the compactified
dimensions. We obtain this map in the explicit form and use it in
formulation of
a technique for generation of the solutions for the theory with arbitrary
value of $d$ starting from the theory with $d=2$. We show that the
$d=2$ theory coincides with the stationary Einstein gravity coupled to the
dilaton field in view of our truncation. Thus, the proposed
generation procedure realizes the general theory extension
\be\label{i1}
{\bf\rm Stationary\,\, Einstein\!-\!Dilaton\,\, Theory}\rightarrow
{\bf\rm Bosonic\,\, String\,\, Theory},
\ee
which allows one to construct a lot of concrete solutions in he low-energy
bosonic string theory. We take the Kerr-NUT solution \cite{EMT}
generalized by the presence of the charged dilaton field and
calculate all the potentials related to this solution and necessary
for our generation procedure. The resulting bosonic string theory
solution describes a non-rotating point-like object with the nontrivial
dilaton and Kalb-Ramond field multipole characteristics. In Conclusion we
discuss the obtained results and give some natural following perspectives.


\section{A model}
\setcounter{equation}{0}
We consider the effective field theory which describes the bosonic string
theory massless modes living in the multidimensional space-time. Let $d+3$,
\,\,$(+ \cdots +;-,++)$ and $X^M$ be the space-time dimensionality, signature
and the coordinates respectively ($M=1,...,d+3$); whereas $\Phi$, $B_{MN}$ and
$G_{MN}$ denote the dilaton, Kalb-Ramond and metric fields. The action
representing a low-energy dynamics of these fields reads \cite{Kir}:
\be\label{m1}
S_{(d+3)}=\int d X\sqrt{-{\rm det}G_{MN}}e^{-\Phi}\left ( R_{d+3}+
\Phi_{,M}\Phi^{,M}-\frac{1}{12}H_{MNK}H^{MNK}
\right ),
\ee
where $H_{MNK}=B_{NK,M}+B_{MN,K}+B_{KM,N}$. In this
article we deal with some consistent truncation of this theory. To
perform it, let us denote $y^m=X^m$ ($m=1,...,d$) and $x^{\mu}=X^{d+\mu}$
($\mu=1,2,3$). Then the truncation can be defined using a combination of
the on-shell consistent restrictions
\be\label{m2}
B_{mn}=G_{m,d+\mu}=B_{d+\mu,d+\nu}=0
\ee
with the following toroidal compactification of the first $d$ dimensions.

Thus, the remaining nontrivial field components depend on the coordinates
$x^{\mu}$ and consist of the quantities $G_{mn}$, $B_{m,d+\mu}$ (we combine
them to the $d\times d$ and $d\times 1$ matrices $G$ and $b_{\mu}$
respectively), and also of the fields $G_{d+\mu,d+\nu}$ and $\Phi$.
Let us also put (\cite{Sen3})
\be\label{m3}
h_{\mu\nu}&=&e^{-2\phi}G_{d+\mu,d+\nu},
\nonumber\\
\phi&=&\Phi-{\rm ln}\sqrt{-{\rm det}G};
\ee
below all the $3$-dimensional operations will be related to the $3$-metric
$h_{\mu\nu}$ and its inverse. Then, let us introduce on shell of the motion
equations the $d\times 1$ pseudo-scalar field $v$ by help of the relation
\cite{Sen3}, \cite{HK1}
\be\label{m4}
\nabla v=e^{-2\phi}G^{-1}\nabla\times\vec b,
\ee
where $(\vec b)_{\mu}=b_{\mu}$. Then the dynamics of the $3$-fields $G$, $\phi$,
$v$ and $h_{\mu\nu}$ is given by the equations \cite{HK}
\be\label{m5}
\nabla\vec J&=&0,
\nonumber\\
R_{3\,\,\mu\nu}&=&\frac{1}{4}{\rm Tr}\left ( J_{\mu}J_{\nu}\right ),
\ee
where the Ricci tensor $R_{3\,\,\mu\nu}$ corresponds to the 3-metric
$h_{\mu\nu}$, $\vec J=\nabla {\cal G}\,{\cal G}^{-1}$ and
\be\label{m6}
{\cal G}=\left (\ba{ccc}
-e^{-2\phi}+v^TGv&\,\,&v^TG\cr
Gv&\,\,&G
\ea\right ).
\ee
Also it is possible to introduce the vector matrix $\vec\Omega$, satisfying
the relation
\be\label{m7}
\nabla\times\vec\Omega=\vec J,
\ee
on shell of the system (\ref{m5}).

Now let us suppose that we have found a solution of the effective
$3$-dimensional problem under consideration, i.e. the quantities ${\cal G}$,
$\vec\Omega$ and $h_{\mu\nu}$ satisfying Eq. (\ref{m5}) and
(\ref{m7}). Then, as it follows from Eq. (\ref{m6}),
\be\label{m8}
G&=&{\cal G}_{22},
\nonumber\\
\quad e^{2\phi}&=&-\left ({\cal G}^{-1}\right )_{11},
\ee
where the indexes correspond to the matrix blocks in Eq. (\ref{m6}). This
completely defines the $(d+3)$-dimensional line element:
\be\label{m9}
ds_{d+3}^2&=&dy^TGdy+e^{2\phi}ds_3^2,
\ee
where $y$ is the $d\times 1$ column with the components $y^m$ and
$ds_3^2=h_{\mu\nu}dx^{\mu}dx^{\nu}$.
Then, as it is easy to prove, the remaining non-trivial $(d+3)$-dimensional
fields are:
\be\label{m10}
B_{m,d+\mu}&=&-\left (\vec\Omega_{21}\right )_{m\mu},
\nonumber\\
e^{\Phi}&=&\sqrt{{\rm det}{\cal G}}\,e^{2\phi}.
\ee
Eqs. (\ref{m8})-(\ref{m10}) translate a solution of the effective
$3$-dimensional system to the solution of the bosonic string theory
(\ref{m1}).


\section{A projection symmetry}
\setcounter{equation}{0}
The simplest solution of the low-energy bosonic string theory (\ref{m1})
satisfying the restrictions (\ref{m2}) describes the empty $(d+3)$-dimensional
space-time. For this solution the corresponding $3$-dimensional quantities are
${\cal G}={\cal G}_0={\rm diag}(-1,-1;1\cdots 1)$, $\vec\Omega=0$ and
$h_{\mu\nu}=\delta_{\mu\nu}$. In this section we construct some symmetry
extension ${\cal G}_0\rightarrow{\cal G}$ of this trivial solution to the field
of non-trivial ones of the form
\be\label{p1}
{\cal G}={\cal S}{\cal G}_0,
\ee
where ${\cal S}$ is the symmetry operator, and calculate corresponding
quantities $\vec\Omega$ and $h_{\mu\nu}$. Our consideration, as well as the
final results will have a form of projection of the original theory living in
$d+3$ dimensions to the $(d^{'}+3)$-dimensional one with $d^{'}\leq d$.

Let us start with a set of the $(d+1)\times 1$ constant columns
$\{C_k\}$ \,\, ($k=1,...,{\cal D}$) and consider the following ${\cal D}^2$
constant matrices:
\be\label{p2}
\Pi_{kl}=C_kC_l^T{\cal G}_0.
\ee
Their mutual products read:
\be\label{p3}
\Pi_{kl}\Pi_{mn}=\kappa_{lm}\Pi_{kn},
\ee
where
\be\label{p3'}
\kappa_{lm}={\rm Tr}\,\Pi_{ml}=C^T_m{\cal G}_0C_l
\ee
is the set of the related constants. We name the matrices $\Pi_{kl}$
``projectors'' in view of Eq. (\ref{p3}): from this it follows that
$\Pi_{kl}^2\sim\Pi_{kl}$, i.e. the essentially projection property. From
Eq. (\ref{p1}) one obtains another identity:
\be\label{p4}
\Pi^T_{kl}={\cal G}_0\Pi_{lk}{\cal G}_0;
\ee
it relates the matrix and index symmetric properties of the
projectors.

Now let us consider the index-free matrix
\be\label{p5}
{\Gamma}=\xi_{kl}\Pi_{kl},
\ee
where the non-matrix quantities $\xi_{kl}$ are introduced and a summation
over the repeating indexes is understood. Using Eqs. (\ref{p3}), (\ref{p4})
it is easy to prove that
\be\label{p7}
{\Gamma}^n=\xi^{(n)}_{kl}\Pi_{kl},
\ee
where
\be\label{p8}
\xi^{(n)}=(\xi\kappa)^n\kappa^{-1},
\ee
and the quantities $\xi_{kl}$ and $\kappa_{kl}$ are naturally embedded to
the ${\cal D}\times {\cal D}$ matrices $\xi$ and $\kappa$ respectively; we
also suppose that ${\rm det}\,\kappa\neq 0$. We define the symmetry
operator as
\be\label{p9}
{\cal S}={\rm exp}\,{\Gamma};
\ee
from Eqs. (\ref{p7}), (\ref{p8}) it follows that
\be\label{p10}
{\cal S}=1+\left [(s-1)\kappa^{-1}\right ]_{kl}\Pi_{kl},
\ee
where $s={\rm exp}\,(\xi\kappa)$. Thus, in the our construction $\xi$ is the
$3$-dimensional matrix field. We need in knowledge of its algebraic and
dynamic properties for complete definition of the symmetry operator
${\cal S}$.

The algebraic property immediately follows from symmetry of the matrix
${\cal G}$ (see Eq. (\ref{m6})): this symmetry is guaranteed if
${\cal S}^T={\cal G}_0\,{\cal S}\,{\cal G}_0$,
i.e. if
\be\label{p11}
s^T=\kappa s\kappa^{-1}
\ee
in view of Eq. (\ref{p10}). From Eq. (\ref{p11}) it also follows that
$\xi^T=\xi$. Then, the dynamical property of the matrix field $s$ can be
obtained from the supposition that the matrix ${\cal G}$ defined by Eqs.
(\ref{p1}), (\ref{p10}) identically satisfies the motion equations
(\ref{m5}). After some algebraical manipulations one concludes that them
actually satisfied if
\be\label{p12}
\nabla\vec j_s&=&0,
\nonumber\\
R_{3\,\,\mu\nu}&=&\frac{1}{4}{\rm Tr}\,\left ( j_{s\,\mu}j_{s\,\nu}\right ),
\ee
where $\vec j_s=\nabla s\,s^{-1}$ and the ${\cal G}$- and $s$-matrix currents
are related as $\vec J=(\vec j_s\kappa^{-1})_{kl}\Pi_{kl}$. Finally, Eqs.
(\ref{p11}) and (\ref{p12}) give the remaining part of information about the
symmetry operator: it is defined in terms of a solution of the dynamical
problem (\ref{p12}) restricted by Eq. (\ref{p11}). Also from Eq. (\ref{p10}) it
follows that the trivial extension in Eq. (\ref{p1}) takes place if $s=1$.

For us it will be useful to reformulate the problem for the matrix field $s$
in the form similar to the one for the matrix ${\cal G}$. To do it, let us
represent the symmetric matrix $\kappa$ in the form of
\be\label{p13}
\kappa=\tau^Tg_0\tau,
\ee
where $g_0$ is the signature matrix for $\kappa$. Then, let us introduce
the following matrix field $g$:
\be\label{p14}
g=g_0\tau s\tau^{-1}.
\ee
It is easy to prove, that the field $g$ satisfies the dynamical problem
of the same type as the `physical' matrix ${\cal G}$:
\be\label{p15}
\nabla\vec j&=&0,
\nonumber\\
R_{3\,\,\mu\nu}&=&\frac{1}{4}{\rm Tr}\, \left ( j_{\mu}j_{\nu}\right ),
\ee
where $\vec j=\nabla g\,g^{-1}$; whereas the algebraical restriction
coincides with the symmetric property:
\be\label{p16}
g^T=g.
\ee
For the completeness of the $g$-formalism let us also introduce on-shell
the vector matrix field $\vec\omega$ as
\be\label{p17}
\nabla\times\vec\omega=\vec j.
\ee
Finally, we have reduced the ${\cal G}$-problem to the $g$-problem of the
same type. From Eq. (\ref{p14}) it follows that ${\cal G}={\cal G}_0$
if $g=g_0$. This means, in particular, that the map $g\rightarrow {\cal G}$
preserves the asymptotic flatness property of the fields if we consider
the $g$-theory in the same manner as the theory related to the `physical'
matrix  ${\cal G}$.

At the end of this section let us establish an explicit form of the relation
between the $g$- and ${\cal G}$-theories. First of all, it is convenient
to introduce the 'normalized' columns $C_{0\,\,k}=
\left [ (\tau^T)^{-1}\right ]_{kl}C_l$. Second, let us combine the columns
$C_{0\,\,k}$ to the single $(d+1)\times{\cal D}$ matrix $C_0$ in the natural
way (so that $C_{0\,\,k}$ is the $k$-th column of $C_0$). Then from Eq.
(\ref{p3'}) one has:
\be\label{p18}
{\cal C}^T{\cal G}_0{\cal C}=g_0.
\ee
The resulting expressions for the $g\rightarrow{\cal G}$ map is remarkable
$\kappa$-independent; after some algebra one concludes that
\be\label{p19}
{\cal G}&=&{\cal G}_0+{\cal C}\left ( g_0gg_0-g_0\right ){\cal C}^T,
\nonumber\\
{\cal G}^{-1}&=&{\cal G}_0+{\cal C}\left ( g^{-1}-g_0\right ){\cal C}^T,
\nonumber\\
\vec\Omega&=&{\cal C}g_0\vec\omega{\cal C}^T{\cal G}_0,
\ee
and also
\be\label{p20}
{\rm det}\,{\cal G}={\rm det}\,g_0\,\,{\rm det}\,g.
\ee
Using the simple algebraical analysis, it is possible to show that Eq.
(\ref{p18}) is compatible only if ${\cal D}\leq d+1$. In the case of exact
equality one deals with application of the hidden symmetries related to our
truncated bosonic string gravity model. Here the number ${\cal D}$
has a clear sense because $d={\cal D}-1$ is the number of toroidally
compactified dimensions. It is possible to generalize this interpretation to
the case of ${\cal D}<d+1$: here we obtain a new type of symmetry extension
of the theory with $d^{'}={\cal D}-1$ compactified dimensions to the theory
with $d>d^{'}$ ones. It is easy to see that this new symmetry map cannot be
generalized itself by the help of isometries acting inside of the $g$- and
${\cal G}$-theories if these isometries preserve a truncation conditions
(\ref{m2}) for the both theories. Such generalizations act as the
re-parameterizations of the matrix $C_0$ and preserve Eq. (\ref{p18}). At the
end of this section let us note that the constructed new symmetry map has the
explicitly projection type.


\section{A generation procedure}
\setcounter{equation}{0}

Now it is possible to apply the decoding relations (\ref{m8})-(\ref{m10})
to the projection scheme given in Eqs. (\ref{p18})-(\ref{p20}).
Let us introduce the
$d\times d$ matrix $G_0={\rm diag}\,\,(-1;1,\cdots 1)$, which denotes the
trivial $G$-value; let us also parameterize the projection matrix $C_0$ as
\be\label{g1}
C_0=
\left (\ba{c}
{\cal A}\cr
{\cal B}
\ea\right ),
\ee
where the matrix blocks ${\cal A}$ and ${\cal B}$ are of the dimensions
$1\times {\cal D}$ and $d\times {\cal D}$ respectively. Then from  Eq.
(\ref{p18}) it follows that
\be\label{g2}
{\cal B}^TG_0{\cal B}-{\cal A}^T{\cal A}=g_0.
\ee
Then, application of Eqs. (\ref{m8}), (\ref{m9}) gives the explicit form
of the $(d+3)$-dimensional line element,
\be\label{g3}
ds_{d+3}^2=dy^T\left [ G_0+{\cal B}(g_0^{-1}gg_0-g_0){\cal B}^T\right ] dy+
\left [ 1-{\cal A}(g^{-1}-g_0){\cal A}^T\right ] ds_3^2,
\ee
whereas by the help of Eq. (\ref{m10}) one obtains the result for the
$(d+3)$-dimensional dilaton and the non-trivial components of Kalb-Ramond
fields:
\be\label{g4}
&&e^{\Phi}=\sqrt{{\rm det}\left ( g_0g\right )}\left [
1-{\cal A}(g^{-1}-g_0){\cal A}^T\right ]
\nonumber\\
&&B_{m,d+\mu}=\left ({\cal B}\omega_{\mu}g_0{\cal A}^T\right )_m.
\ee
Eqs. (\ref{g3})-(\ref{g4}) define a solution of the bosonic
string gravity system (\ref{m1}) truncated accordingly Eq. (\ref{m2})
in terms of the constant matrices ${\cal A}$ and ${\cal B}$ restricted
by Eq. (\ref{g2}) and the matrix fields $g$ and $\vec\omega$ satisfying
Eqs. (\ref{p15})-(\ref{p17}). This result takes place for the arbitrary
${\cal D}\leq d+1$.

Now let us consider the special case of ${\cal D}=2$ and $d\geq 2$, i.e.
the essentially projection case
(the bosonic string gravity with $d=1$ was considered in \cite{Bak});
let us also put $g_0=1$ for definiteness.
Then the general solution of the algebraic equation (\ref{g2}) reads:
\be\label{g5}
{\cal A}=(a_1\,\,a_2),
\qquad
{\cal B}=
\left (\ba{ccc}
b_1&\,\,&b_2\cr
\quad &\,\,&\quad\cr
\sqrt{1+a_1^2+b_1^2}\,\,n_1&\,\,&
\sqrt{1+a_2^2+b_2^2}\,\,n_2
\ea\right ),
\ee
where $a_1, a_2;\, b_1, b_2$ are the arbitrary parameters, whereas
$n_1$ and $n_2$ are the $(d-1)\times 1$ columns. These columns satisfy the
restrictions
\be\label{g6}
&&n_1^Tn_1=n_2^Tn_2=1,
\nonumber\\
&&n_1^Tn_2=\frac{a_1a_2+b_1b_2}{\sqrt{1+a_1^2+b_1^2}\sqrt{1+a_2^2+b_2^2}}
\ee
and can be considered as the pair of unimodular $(d-1)$-dimensional
vectors with some angle $\delta$ between them defined by the relation
$\delta=\arccos{(n_1^Tn_2)}$. For $d>2$ the parameters $a_1, a_2, b_1, b_2$
are actually free because
$(a_1a_2+b_1b_2)/\sqrt{(1+a_1^2+b_1^2)(1+a_2^2+b_2^2)}\leq
(a_1a_2+b_1b_2)/\sqrt{(a_1^2+b_1^2)(a_2^2+b_2^2)}\leq 1$; this
makes a definition of the angle $\delta$ through these parameters correct.
In the case of $d=2$ one has $\delta=0,\,\pi$, and Eq. (\ref{g6}) can also
be satisfied for some limit values of the discussing parameters.

Then, let us parameterize the dynamical matrices $g$ and $\vec\omega$
in the following way:
\be\label{g7}
g=e^{\psi}
\Psi_4^{-1}\left (\ba{ccc}
\Psi_1&\,\,&\Psi_3\cr
\Psi_3&\,\,&\Psi_2
\ea\right ), \qquad
\vec\omega=
\left (\ba{ccc}
\vec\omega_1&\,\,&\vec\omega_3\cr
\vec\omega_4&\,\,&\vec\omega_2
\ea\right ),
\ee
where $\Psi_4^2=\Psi_1\Psi_2-\Psi_3^2$ and the scalar and vector
components are related accordingly Eq. (\ref{p17}). Let us denote $t=y^1$
(the time coordinate) and $Y^r=y^{r+1}$\,\,(the extra dimension coordinates,\,
$r=1,...,d-1$). Then for the $(d+3)$-dimensional line element
one obtains that
\be\label{g8}
ds_{d+3}^2=-F(dt-K^TdY)^2+dY^THdY+Eds_3^2,
\ee.
where the quantities $F$, $K$, $H$ and $E$ read:
\be\label{g9}
&&F=1+b_1^2+b_2^2-e^{\psi}\Psi_4^{-1}\left ( b_1^2\Psi_1
+b_2^2\Psi_2+2b_1b_2\Psi_3\right ),
\nonumber\\
&&K=F^{-1}\left\{
\sqrt{1+a_1^2+b_1^2}\left [-b_1+e^{\psi}\Psi_4^{-1}
\left ( b_1\Psi_1+b_2\Psi_3\right )\right ] n_1
+
\sqrt{1+a_2^2+b_2^2}\left [-b_2
\right.\right. \nonumber\\
&&\hskip 11mm \left.\left.
+e^{\psi}\Psi_4^{-1}
\left ( b_2\Psi_2+b_1\Psi_3\right )\right ] n_2
\right\},
\nonumber\\
&&H=1+
(1+a_1^2+b_1^2)\left\{-1+e^{\psi}\Psi_4^{-1}\Psi_1+F^{-1}
\left [ -b_1+e^{\psi}\Psi_4^{-1}\left ( b_1\Psi_1+b_2\Psi_3\right )\right ]^2
\right\}n_1n_1^T
\nonumber\\
&&\hskip 13mm+(1+a_2^2+b_2^2)\left\{-1+e^{\psi}\Psi_4^{-1}\Psi_2+F^{-1}
\left [ -b_2+e^{\psi}\Psi_4^{-1}\left ( b_2\Psi_2+b_1\Psi_3\right )\right ]^2
\right\}n_2n_1^2
\nonumber\\
&&\hskip 13mm+
\sqrt{(1+a_1^2+b_1^2)(1+a_2^2+b_2^2)}\left\{e^{\psi}\Psi_4^{-1}\Psi_3+
F^{-1}
\left [ -b_1+e^{\psi}\Psi_4^{-1}\left ( b_1\Psi_1+b_2\Psi_3\right )\right ]
\right.
\nonumber\\
&&\hskip 13mm\times\left.
\left [ -b_2+e^{\psi}\Psi_4^{-1}\left ( b_2\Psi_2+b_1\Psi_3\right )\right ]
\right\}\left ( n_1n_2^T+n_2n_1^T\right ),
\nonumber\\
&&E=1+a_1^2+a_2^2-e^{-\psi}\Psi_4^{-1}\left (a_2^2\Psi_1+
a_1^2\Psi_2-2a_1a_2\Psi_3\right );
\ee
whereas the non-zero components of the $(d+3)$-dimensional matter fields
are
\be\label{g10}
&&e^{\Phi}=e^{\psi}\left (1+a_1^2+a_2^2\right )-\Psi_4^{-1}\left (a_2^2\Psi_1+
a_1^2\Psi_2-2a_1a_2\Psi_3\right ),
\nonumber\\
&&B_{t,d+3}=(a_1\omega_1+a_2\omega_3)b_1+(a_2\omega_2+a_1\omega_4)b_2,
\nonumber\\
&&B_{r,d+3}=\sqrt{1+a_1^2+b_1^2}(a_1\omega_1+a_2\omega_3)n_{1\,r}
+\sqrt{1+a_2^2+b_2^2}(a_2\omega_2+a_1\omega_4)n_{2\, r}.
\ee
Eqs. (\ref{g8})-(\ref{g10}) define the generation procedure completely;
the generation parameters are restricted accordingly Eq. (\ref{g6}). It is
easy to see that this procedure possesses a wide group of symmetry
transformations itself. Actually, the restrictions (\ref{g6}) on the columns
$n_1$, $n_2$ preserve the form if one rotates the column pair `as the solid
body' in the effective $(d-1)$-dimensional parametric space. Then, it is
possible to combine the parameters $a_1, a_2, b_1, b_2$ to the $2$-dimensional
vectors $(a_1,\,b_1)$ and $(a_2,\,b_2)$. It is clear that the right hand side
of the second equation in (\ref{g10}) is invariant in respect to rotations in
this space. Finally, the full parametric symmetry group of the our generation
procedure is $O(d-1)\times O(2)$. This symmetry group acts as the group
of re-parameterizations of the numeric parameters of the solution; in
applications it can be convenient to use it for the `gauge fixation' or,
equivalently, for the concrete choice of frame in the effective parametric
space of the solution.


\section{A solution}
\setcounter{equation}{0}

Let us consider the generation base, i.e., the theory related to the dynamical
matrices $g$ and $\vec\omega$. Let us put
\be\label{s1}
g=e^{\psi}\hat g;
\ee
then from Eq. (\ref{g7}) it follows that ${\rm det}\,\hat g=1$, i.e.
$\hat g\in SL(2,R)/SO(2)$. In terms of the parameterization (\ref{s1}) the
motion equations (\ref{p15}) take the form
\be\label{s2}
\nabla\hat{\vec j}&=&0,
\qquad
\nabla^2\psi=0,
\nonumber\\
R_{3\,\,\mu\nu}&=&\frac{1}{4}{\rm Tr}\,\left ( \hat j_{\mu}\hat j_{\nu}
\right )+\frac{1}{2}\psi_{,\mu}\psi_{,\nu},
\ee
where $\hat{\vec j}=\nabla\hat g\,\,\hat g^{-1}$. Then, for the matrix
$\vec\omega$ one has the following decomposition:
\be\label{s3}
\vec\omega=\hat{\vec\omega}+\tilde{\vec\omega},
\ee
where $\hat{\vec\omega}=\hat{\vec j}$ and $\tilde{\vec\omega}=\nabla\psi$.

The system under consideration is equivalent on-shell to the stationary
$4$-dimensional Einstein-dilaton theory,
here the matrix $\hat g$ is constructed from the pure Einstein variables,
whereas $\psi$ denotes the effective dilaton field. This interpretation is
alternative to the one given above: there the model related to the matrix
field $g$ was the $4$-dimensional bosonic string theory truncated accordingly
Eq. (\ref{m2}). In framework of the new interpretation the variables
$\Psi_{\alpha}$,\,\,$\alpha=1,...,4$ are related to the Ernst
representation of the stationary Einstein gravity as
\be\label{s4}
{\cal E}=\Psi_1^{-1}\left ( \Psi_4+i\Psi_3\right ),
\ee
where ${\cal E}$ is the Ernst potential \cite{Ernst}. Our goal is to use a
solution space of the General Relativity for calculation of the functions
$\Psi_{\alpha}$ and $\vec\omega_{\alpha}$ which define the
$(d+3)$-dimensional fields of the bosonic string theory accordingly Eqs.
(\ref{g8})-(\ref{g10}). To
do it, let us suppose that the problem (\ref{s2}) is axisymmetric, i.e.
the fields are $\varphi$-independent in the canonic Weil coordinates $\rho, z,
\varphi$ \cite{EMT}. In these coordinates the $3$-dimensional line element
reads:
\be\label{s5}
ds_3^2=\exp{(\gamma)}\left ( d\rho^2+dz^2\right )+\rho^2d\varphi^2;
\ee
the $\hat g$- and $\psi$-equations in (\ref{s2}) are $\gamma$-independent and,
thus, they defined over the flat $3$-space. Then,
\be\label{s6}
\gamma=\hat\gamma+\tilde\gamma,
\ee
where the functions $\hat\gamma$ and $\tilde\gamma$ are defined by the
Einstein part of Eq. (\ref{s2}) for the case of trivial value of $\psi$ and
$\hat g$ respectively. In this scheme the $\tilde\gamma$-problem reads:
\be\label{s7}
\tilde\gamma_{,z}&=&\psi_{,\rho}\psi_{,z}\nonumber\\
\tilde\gamma_{,\rho}&=&\frac{1}{2}(\psi_{,\rho}^2-\psi_{,z}^2).
\ee
Let us also note that in the axisymmetric case the vector quantities
$\hat{\vec\omega}$ and $\tilde{\vec\omega}$ have the non-zero
$\varphi$-components only. Finally, to obtain the base for generation
accordingly Eqs. (\ref{g8})-(\ref{g10}), one must take a solution of the
Einstein problem and to calculate the potentials $\Psi_{\alpha},\hat\omega_
{\alpha,\,\varphi},\hat\gamma$ and also a solution $\psi$ of the
Laplace equation together with the related functions $\tilde\omega_{\alpha},
\tilde\gamma$. As it was shown above, to obtain asymptotically flat
$(d+3)$-dimensional solution one must start from solution of the effective
Einstein-dilaton theory possessing the same property.

Below we give the expressions for the all necessary quantities in terms of
the coordinates $R,\theta,\varphi$ which relate to the canonical Weil ones
as
\be\label{s8}
\rho &=& \sqrt{R^2+\alpha^2}\sin^2\theta,
\nonumber\\
z &=& R\cos\theta.
\ee
We consider the class of Kerr-NUT solutions of Einstein theory coupled to
the generalized Coulomb solution for the effective dilaton field. The
corresponding $\Psi_{\alpha}$-potentials are:
\be\label{s9}
\Psi_1&=&(R+m_1)^2+(a\cos\theta-m_2)^2,
\nonumber\\
\Psi_2&=&(R-m_1)^2+(a\cos\theta+m_2)^2,
\nonumber\\
\Psi_3&=&-2(m_2R+m_1a\cos\theta),
\nonumber\\
\Psi_4&=&R^2+a^2\cos^2\theta-|m|^2;
\ee
whereas the potential $\psi$ reads:
\be\label{s10}
\psi=\frac{2(q_1-q_2\alpha\cos\theta)}{R^2+\alpha^2\cos^2\theta}.
\ee
Here $m=m_1+im_2$ and $q=q_1+iq_2$ are the arbitrary complex constants, $a$
is the real parameter and $\alpha=\sqrt{a^2-|m|^2}$. Then, the potentials
$\omega_{\alpha,\,\varphi}$ are:
\be\label{s11}
\omega_{1\,\varphi}=2(q_1+m_1)\cos{\theta}&+&2\sin{\theta}^2\left [
\frac{\alpha (q_2R+q_1\alpha\cos{\theta})}{R^2+\alpha^2\cos{\theta}^2}+
\frac{a(m_2R+m_1a\cos{\theta})}{\Psi_4}
\right ],
\nonumber\\
\omega_{2\,\varphi}=2(q_1-m_1)\cos{\theta}&+&2\sin{\theta}^2\left [
\frac{\alpha (q_2R+q_1\alpha\cos{\theta})}{R^2+\alpha^2\cos{\theta}^2}-
\frac{a(m_2R+m_1a\cos{\theta})}{\Psi_4}
\right ],
\nonumber\\
\omega_{3\,\varphi}=-2m_2\cos{\theta}&+&2\sin{\theta}^2
\frac{a(m_1R-m_2a\cos{\theta}+|m|^2)}{\Psi_4},
\nonumber\\
\omega_{4\,\varphi}=-2m_2\cos{\theta}&+&2\sin{\theta}^2
\frac{a(m_1R-m_2a\cos{\theta}-|m|^2)}{\Psi_4}.
\ee
Finally, the $3$-metric for this solution is:
\be\label{s12}
ds^2_3&=&
{\rm exp}\left (-\frac{\sin^2\theta}{2}
\left\{
\frac{q_1^2+q_2^2}{R^2+\alpha^2\cos^2\theta}+
\frac{R^2+\alpha^2}{(R^2+\alpha^2\cos^2\theta)^4}
\left [
(q_1^2-q_2^2)
(R^4-6R^2\alpha^2\cos^2\theta
\right. \right. \right.
\nonumber\\
&+&\left. \left. \left.
\alpha^4\cos^4\theta)-
8q_1q_2R\alpha (R^2-\alpha^2\cos^2\theta)\cos\theta
\right ]
\right\}\right )
\Psi_0
\left ( \frac{dR^2}{R^2+\alpha^2\cos^2\theta}+d\theta^2\right )
\nonumber\\
&+&
(R^2+\alpha^2\cos^2\theta)\sin^2\theta d\varphi^2.
\ee
Eqs. (\ref{g8})-(\ref{g10}) and (\ref{s9})-(\ref{s12})
completely define our class of solutions for the
$(d+3)$-dimensional low energy bosonic string theory. This
class depends on the set of $2d+7$ parameters restricted by
$3$ relations, i.e. the constructed solution class has $2d+4$
parametric degrees of freedom.

At the end of this section let us discuss our choice of
solution for the generation base, i.e. for the effective
Einstein-dilaton theory. Our statement is that for the solution
taken the Einstein and dilaton parts have the same formal
nature. Actually, let us calculate the potential $\zeta=
(1-{\cal E})/(1+{\cal E})$ (\cite{Maz}) for the Kerr-NUT
solution; the result reads:
 \be\label{s13}
{\zeta}=\frac{m}{R-ia\cos\theta}.
\ee
Then, let us consider the complex solution of the axisymmetric
Laplace equation given by the function $\xi$,
\be\label{s14}
\xi=\frac{q}{R-i\alpha\cos\theta};
\ee
it is seen that $\zeta\rightarrow\xi$ if $m\rightarrow q$ and
$a\rightarrow\alpha$. Then our statement is related to the fact that
$\psi=\xi+\bar\xi$.

\section*{Acknowledgments}
This work was supported by RFBR grant ${\rm N^{0}}
\,\, 00\,02\,17135$.


\section{Conclusion}
\setcounter{equation}{0}

In this article we have established a general form of the relation
between the solution spaces of the effective field theories resulting
from the toroidal compactification of the low-energy bosonic string
theory from the diverse to three dimensions. Using this relation, we
have developed a technique for generation of new solutions for the
theory living in the space-time of the arbitrary dimensionality
starting from the stationary $4$-dimensional Einstein-dilaton gravity.
We have calculated the special solution which describes a new bosonic
string theory extension of the Kerr-NUT solution modified by the presence
of the dilaton field of the Coulomb type. This concrete solution is not
of the form known in the literature: it is static from the point of view
of the physical $3$-dimensional space. Nevertheless, this solution
possesses the rotational properties, but its rotation related to
the extra dimensions. The direction of a rotational axis in the extra
dimensions  is defined by the pair of the unimodular vectors $n_1$ and $n_2$,
and depends on the position in the physical $3$-space. The constructed
solution is charged and possesses the peculiarities of the Dirac type for the
Kalb-Ramond field components. These peculiarities can be removed using some special
choice of the free solution parameters; we leave the following physical
analysis and discussion to the future.

The perspectives are related to the possible generalizations of the obtained
results to the case of the non-constrained bosonic string gravity and,
moreover, to the case of the gravity model originated from the low-energy
heterotic string theory. The corresponding gravity model generalizes the
one considered above by the presence of some number of the $U(1)$ gauge
fields. In \cite{HK2}, \cite{OK} for such gravity model with the arbitrary
number of the Abelian gauge fields and toroidally compactified to three
dimensions it was developed a formalism which seems really promising for
the solution of the discussing problems. Moreover, there is a hope to
generalize the projection formalism and the corresponding generation
procedure to the case of arbitrary theory possessing a symmetric space
model representation after the compactification to three dimensions \cite{BGM}.
Also the important perspective is related to formulation of the generation
procedure based on the stationary Einstein-Maxwell theory. Its realization
allows one to extend the widest solution space of the Einstein-Maxwell
theory to the case of weakly studied field theory limits of the superstring
theory. Coming back to the symbolic formula (\ref{i1}), let us represent the
future perspectives in the following form:
\be\label{c1}
{\bf\rm Stationary\,\, Einstein\!-\!Maxwell\,\, Theory}\rightarrow
{\bf\rm Heterotic\,\, String\,\, Theory}.
\ee
Of course, the natural and the nearest application of a such formalism
is the generation of the heterotic string theory extension of
the Kerr-Newman solution with the nontrivial magnetic and NUT charges.
The first step in this direction had been performed in
(\cite{OK1}), where the special case of the $4$-dimensional heterotic string
theory with two Abelian gauge fields was considered. This special case
was considered in the literature in view of its  relation to the $D=N=4$
supergravity (see \cite{Kal}, where the supergravity generalization of
the Israel-Wilson-Perjes solution of Einstein-Maxwell theory was obtained
and also \cite{Sab} for the black hole solution construction in supergravity
using harmonic functions). In (\cite{OK1}) it was shown that
for this special heterotic string gravity model the generation program
can be realized in terms of some single complex $2\times 2$ matrix
potential which transforms linearly under the action of symmetries
preserving the asymptotic flatness property of the fields. For the general
case of the arbitrary original space-time dimensionality and number
of the gauge fields it seems the most natural to use a language of the
real matrix generalization of the $\zeta$-potential developed for the
heterotic string gravity in \cite{OK}. A solution of the discussing problems
is now in progress.


\end{document}